\documentclass[letterpaper]{article} 
\usepackage[]{aaai25}  
\usepackage{times}  
\usepackage{helvet}  
\usepackage{courier}  
\usepackage[hyphens]{url}  
\usepackage{graphicx} 
\usepackage{natbib}  
\usepackage{caption} 
\usepackage{algorithm}
\usepackage{algorithmic}

\usepackage{newfloat}
\usepackage{listings}
\DeclareCaptionStyle{ruled}{labelfont=normalfont,labelsep=colon,strut=off} 
\floatstyle{ruled}
\newfloat{listing}{tb}{lst}{}
\floatname{listing}{Listing}
\usepackage{mathtools}
\usepackage{float}
\usepackage{amsfonts}
\usepackage{amssymb}
\usepackage{subcaption}
\usepackage{sgame}

\title{Towards Attacker Type Agnostic Cyber Defense Agents}
\author{
    Erick Galinkin,
    Emmanouil Pountourakis,
    Spiros Mancoridis
}
\affiliations{
    \textsuperscript{\rm 1}Drexel University\\

    eg657@drexel.edu
}

\begin{document}

\maketitle

\begin{abstract}
With computing now ubiquitous across government, industry, and education, cybersecurity has become a critical component for every organization on the planet. 
Due do this ubiquity of computing, cyber threats have continued to grow year over year, leading to labor shortages and a skills gap in cybersecurity. 
As a result, many cybersecurity product vendors and security organizations have looked to artificial intelligence to shore up their defenses.
This work considers how to characterize attackers and defenders in one approach to the automation of cyber defense -- the application of reinforcement learning.
Specifically, we characterize the types of attackers and defenders in the sense of Bayesian games and, using reinforcement learning, derive empirical findings about how to best train agents that defend against multiple types of attackers.
\end{abstract}
\section{Introduction}
The use of machine learning in cybersecurity has grown substantially as a way to simultaneously increase the speed and reduce the cost of cyber threat detection and response.
One machine learning technique that has seen considerable interest but seemingly little adoption in deployed systems is reinforcement learning. 
While reinforcement learning is notoriously unstable and can be difficult to evaluate, it has also famously seen tremendous success in cases like AlphaGo~\cite{silver2016mastering}.
The translation of cybersecurity problems into a Markov decision process is a key step in applying techniques like reinforcement learning, and frequently overlap with game theoretic framings of these same cybersecurity problems.

Today, security orchestration, automation, and response (SOAR) frameworks are used by cybersecurity practitioners to reduce the human labor associated with responding to cybersecurity incidents.
These SOAR systems can orchestrate disparate tools within workflows to get additional data, automate remediation actions, and often have other, diverse capabilities.
While some SOAR systems have integrated AI/ML capabilities~\cite{kinyua2021ai}, they often lack true automation in the containment and recovery phases of the Identification, Containment, Eradication, Recovery (ICER) cycle. 
Indeed, most SOAR systems rely on expert written \texttt{if-then} rules to take limited actions when some event is triggered.
Our work aims to use deep reinforcement learning to fill this gap.

One difficulty in many reinforcement learning and game theoretic approaches is the notion of ``type'' in the sense of Harsanyi~\cite{harsanyi1967games}.
That is, the games \textit{a priori} fix the parameters of the attacker and defender to specify goals and capabilites.
In reality, cyber defenders observe a wide variety of attacker types whose tactics, techniques, and procedures (TTPs) may overlap considerably even if their objectives differ. 
This work considers two attacker types with substantially divergent objectives: a ransomware attacker who seeks to gain control of 80\% of a target network and an advanced persistent threat actor who seeks to access data on a single high-value node.
We provide detailed definitions of attacker and defender types below.
In this work, we provide an extension of the YAWNING-TITAN~\cite{andrew2022developing} reinforcement learning framework to allow specification of different attacker types for independent learning agents. 
Additionally, we demonstrate that a models trained across adversary types in a self-play setting yields a robust, attacker type agnostic defensive agent.
We further demonstrate that even when defenders have seen only a single type of attacker, the learned policies are transferable, albeit suboptimal, against unseen adversary types.
\section{Background}
Reinforcement learning (RL) has been widely explored in cybersecurity~\cite{nguyen2021deep,adawadkar2022cyber} as a way to automate detection and response.
If we take a graph view of systems we need to defend -- a natural representation for computer networks -- the data regarding systems is necessarily high-dimensional, capturing the states of individual machines and their network connections. 
Reinforcement learning, and deep reinforcement learning in particular, serves to effectively reduce this dimensionality and make learning tractable.
Deep RL has seen many applications in cybersecurity, including defense of cyber-physical systems~\cite{feng2017deep}; phishing detection~\cite{chatterjee2019detecting}; and moving target defense~\cite{li2023robust}.

This work grounds our experiments using a partially observable stochastic Bayesian game, building on the stochastic Bayesian game~\cite{albrecht2013game}.

In this work, we build upon the YAWNING-TITAN (YT) reinforcement learning framework~\cite{andrew2022developing} which uses Proximal Policy Optimization (PPO)~\cite{schulman2017proximal} as its default reinforcement learning algorithm. 
This particular algorithm has been adopted in a number of applications, including cybersecurity~\cite{nguyen2021deep,adawadkar2022cyber,galinkin2024price} as a powerful on-policy deep RL algorithm.
In contrast to standard YT, we use self-play between two independent networks and extend the framework itself to include ``multi-type training''.
Our implementation of self-play also differs from standard YT by incorporating partial observability and noise~\cite{galinkin2023simulation} to reflect the presence of false positive alerts commonly observed in practice and more accurately frame the difference between the true state of the environment and the observed space.

In addition to standard PPO, our work also considers a Hierarchical PPO (HiPPO) algorithm~\cite{Li2020Sub-policy}.
The HiPPO algorithm leverages a high-level and low-level policy network where a higher-level manager algorithm does not take an action in the space, but rather conditions the low-level policy networks.
At some fixed interval, the manager receives a new observation and decides which low-level policy to commit to over the next interval.
In our case, these low-level policies reflect the ``skills'' of detecting and responding to our different attacker types.
\section{Methodology} \label{sec:methods}
We build upon a partially observable stochastic Bayesian game with noise ala~\cite{galinkin2023simulation}.
In contrast with traditional stochastic Bayesian games~\cite{albrecht2013game} which uses Harsanyi-Bellman ad hoc coordination, the partial observability of the game means that there is presently no standard solution concept. 
We consider two attacker types in our assessment.
These attacker types have the same actions available to them and thus, differ in terms of their objectives.
\begin{enumerate}
    \item Ransomware: an attacker who receives a reward for controlling 80\% or more of the target network
    \item APT: an attacker who aims to access information on a single high-value node
\end{enumerate}
We elect to use these attacker types for several reasons. 
First and foremost, ransomware and APT-style attackers are often considered the highest priority threats for many major businesses.
Secondly, these attacker types have distinctly different objectives, making it easy to contrast them against each other.
In each game, a number of nodes are created, one of which is a ``high-value target''.
The ransomware actor is indifferent to whether or not the node they have compromised is that target, and treats each node with an equal value.
By contrast, the APT actor considers every node except for the high-value target to have a value of zero, with all reward confined to exfiltrating data from that single node.

In line with prior work~\cite{galinkin2024price}, we train our agents in an environment with a shared state space but two separate observation spaces. 
This approach allows the attacking agent to learn through play and requires two distinct instances of proximal policy optimization (PPO)~\cite{schulman2017proximal} -- one for the attacker and one for the defender -- to train the agents.
Since the observation spaces differ and the hidden information is key to our methodology, methods like multi-agent DDPG~\cite{lowe2017multi} are not suitable, as these methods will leak hidden information to the other agent.

We train our agents in four different settings:
\begin{enumerate}
    \item Ransomware: The attacking agent has a ransomware objective throughout training
    \item APT: The attacking agent has an APT objective throughout training
    \item Alternating: Two attacking agents -- one Ransomware and one APT -- are instantiated. At each training epoch, one is chosen at random to play against the defender.
    \item Hierarchical: Two attacking agents -- one Ransomware and one APT -- are instantiated. The defender uses a hierarchical PPO model with one high-level and two low-level policies. At each training epoch, an attacking agent is chosen at random to play against the defender.
\end{enumerate}

\subsection{Environment}
The environment is based on a partially observable stochastic Bayesian game with noise~\cite{galinkin2023simulation} that is intended to reflect realistic conditions.
This game setting reflects the fact that neither the attacker nor the defender has full, true knowledge of the full state space, and that the outcomes of their actions are not fully determined -- there is a probabilistic outcome associated with their action succeeding or failing.
The game is defined as a tuple $\Gamma = (S, A, P, R, \Theta)$ where:
\begin{itemize}
    \item $S = \langle V, E \rangle$ is the state space
    \item $A = \{A_{\mathcal{A}}, A_{\mathcal{D}}\}$ is the action space with $A_{\mathcal{A}}, A_{\mathcal{D}}$ representing the attacker's and defender's action spaces, respectively
    \item $P: S \times S \rightarrow [0, 1]$ is the state transition function representing the probability that a compound action $a_t = \langle a_{\mathcal{A}}, a_{\mathcal{D}}\rangle$ in state $s$ at time $t$ will yield some state $s'$ at time $t+1$
    \item $R$ the expected immediate reward of taking action $a$ in state $s$
    \item $\Theta = \{\Theta_{\mathcal{A}}, \Theta_{\mathcal{D}}\}$ the type spaces for attackers and defenders.
\end{itemize}

The state space represents a computer network that the defender is tasked with defending and which the attacker seeks to compromise to achieve their own goal. 
This is modeled as a network graph where each node is a defender-owned computer and each edge is a network connection between two nodes in that graph. 
Each node $v$ is a tuple $(v_{\mathcal{v}}, v_{\alpha}, v_{\delta})$ that defines the true state and the hidden information:
\begin{enumerate}
    \item $v_{p} \in [0, 1]$: The ``vulnerability'' of a particular node -- the probability that an attack will be successful.
    \item $v_{\alpha}$: The true value of whether the node has been compromised by the attacker, visible only to the attacker.
    \item $v_{\delta}$: Defender-visible attribute that indicates whether an alert has been triggered on the node.
\end{enumerate}

In our game, we fix the action spaces $A_{\mathcal{A}}, A_{\mathcal{D}}$ for attacker and defenders, subject to the actions available in YAWNING-TITAN~\cite{andrew2022developing}, such that all attacker and defender types share a relevant action space.
$A_{\mathcal{A}}$ is comprised of five actions:
\begin{itemize}
    \item Basic Attack: Compromise and make accessible some adjacent $v \in V$ with probability given by $v_{v}$, the ``vulnerability'' of the particular node.
    \item 0-day Attack: Compromise and make accessible some adjacent $v \in V$ even if $v_{v} = 0$.
    \item Move: Move from some compromised $v \in V$ to another accessible $v' \in V$
    \item Do Nothing: Take no action
    \item Execute: End the game and realize rewards for all compromised $v \in V$
\end{itemize}
$A_{\mathcal{D}}$ is comprised of seven actions
\begin{itemize}
    \item Reduce Vulnerability: For some $v \in V$, slightly decrease the probability, $p$ that a basic attack will be successful
    \item Make Node Safe: For some $v \in V$, reduce the probability that a basic attack will be successful to 0.01
    \item Restore Node: For some $v \in V$, reset the node to its initial, uncompromised state
    \item Scan: With some probability, detect the true compromised status of each $v \in V$
    \item Isolate: For some $v \in V$, remove all $e \in E$ connected to it
    \item Reconnect: For some $v \in V$, restore all $e \in E$ that have been disconnected
    \item Do Nothing: Take no action
\end{itemize}
The costs and rewards that specify $R_{\mathcal{A}}, R_{\mathcal{D}}$ are defined by the player's type, detailed below.
Since our attackers and defenders leverage reinforcement learning, the actual function computing $R$ is learned by the ``critic'' value function of the agent.

\subsection{Attacker Type Definition} \label{sec:attacker}
Using the partially observable stochastic Bayesian game as the basis of our analysis, we must consider the type, in the sense of Harsanyi~\cite{harsanyi1967games,harsanyi1968games}, of our attacker and defender.
In this setting, the type, $\theta_{\mathcal{A}}$ of an attacker is drawn from the set of all types $\Theta_{\mathcal{A}}$.
This type is uniquely defined by the objective of the attacker, characterized by their reward function.

\begin{figure}[h!]
    \centering
    \includegraphics[width=0.4\textwidth]{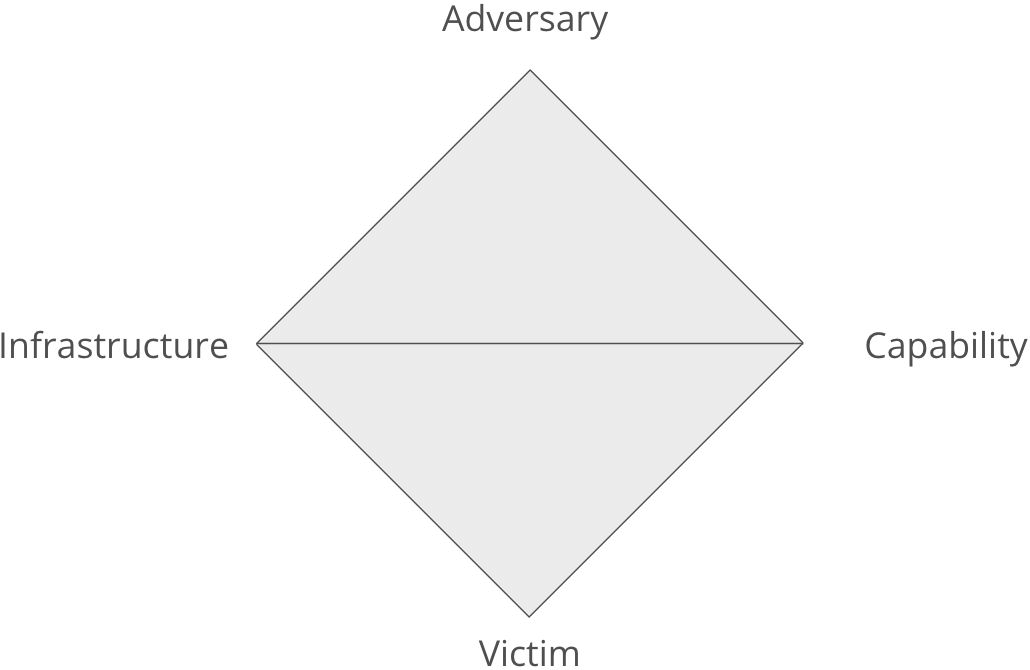}
    \caption{The Diamond Model of intrusion analysis: Adversary, Victim, Capability, and Infrastructure. Edges represent the relationships between features that can be used to analyze and discover malicious activity.}
    \label{fig:diamond_model}
\end{figure}

In cybersecurity, threat actor attribution is frequently used and forms a natural analogy with types.
However, attack attribution is a notoriously difficult practice~\cite{perry2019no} and is one reason why so much of threat intelligence relies on private companies selling this information to defenders.
There are many models for performing attribution of threat activity to a particular group, the most prominent of which is the diamond model of intrusion analysis~\cite{caltagirone2013diamond} that contains four core features: adversary, capability, infrastructure, and victim, pictured in Figure~\ref{fig:diamond_model}, and links those features together to discover knowledge of malicious activity.
The diamond model also includes meta-features like timestamp, phase, result, direction, methodology, and resources. 
In essence, threat intelligence asks and aims to answer the question: ``who is conducting the attack, how is it being conducted, and what do they want?'' 
A full discourse on the diamond model is beyond the scope of this paper but the outcome of this type of analysis is typically an identification of an intrusion set or threat actor group that has a natural analogy to an attacker's type.

This attacker type mapping may be extremely granular \textit{e.g.}, identifying activity associated with a particular military unit ala Mandiant's pioneering report on APT 1~\cite{mandiant2013apt1}.
In many cases, identification is far more coarse -- simply identifying that the threat actor group is some kind of advanced persistent threat or cybercrime group.
In this work, we take a coarse view of attacker types.

\subsection{Defender Type Definition} \label{sec:defender}
While attacker types have a clear parallel to threat actor groups, defender types have largely remained ill-defined.
Our defender type should capture attributes of how a defensive environment is set up -- the detection of malicious activity and potential false positives. 
It should also capture the ``maturity'' of the organization as it represents their ability to respond to threats.
In our setting, we consider a number of parameters that contribute to the optimal strategy of a defensive agent -- specifically:
\begin{itemize}
    \item The probability of detecting an attack $p$
    \item The false positive rate of detection $q$
    \item The cost $c$ associated with defender actions -- a proxy for the organization's maturity
    \item The payoff function $u$ associated with preventing or failing to prevent an attack
\end{itemize}
Since defenders have the same objective -- maintaining the confidentiality, integrity, and availability of a system -- and will learn a strategy parameterized by the above, we say that a defender's type $\theta_{\mathcal{D}}$ consists of the above, plus their learned strategy $\pi_{\mathcal{D}}$.

In contrast with attackers, defenders lack a taxonomization and classification. 
To this end, we suggest characterizing both $p$ and $q$ -- which are typically positively correlated -- both individually and in relation to each other. 
In lieu of a proper taxonomy, we offer some example archetypes below.

\begin{itemize}
    \item Cautious: $1.0 > p > 0.7$, $0.3 > q > 0.2$ -- An organization generally accepting of false positives. Typifies an organization with a large security budget or some managed security service providers. Note that in general, a false positive rate exceeding 30\% is impractical to manage for nearly any organization.
    \item Balanced: $0.7 > p > 0.5$, $0.2 > q > 0.1$ -- Organization with a somewhat constrained security budget who cannot afford to respond to large numbers of false positives. Typifies organizations like a medium-sized business.
    \item Constrained: $0.5 > p > 0.3$, $0.1 > q > 0.0$ -- Organizations with a limited security budget who can afford only to respond to a limited number of threats. Typifies organizations like K-12 education or small businesses.
\end{itemize}

Throughout this work, we fix our defender type and assume that we are operating with a balanced defender who has values of $p = 0.6$ and $q = 0.1$, consistent with prior work~\cite{galinkin2023simulation}. 
That is, we assume the defender has configured their tooling such that they correctly detect 60\% of attacker activity but mischaracterize benign activity as malicious 10\% of the time.
Additionally, we define the defender's utility $u = r - \sum_{t}^T c_t$ where $r$ is $0$ if the defender fails to prevent the attack and $5000$ for successfully eliminating the attacker; $T$ is the total length of the episode; and $c_t$ is the cost of the action taken at timestep $t$.
We note that the reward value for winning is derived by choosing the value of the highest cost action available to the defender ($10$) and multiplying it by the maximum number of possible timesteps per round ($500$).
Modifying these settings impacts the learned policy and should assume realistic values that may be derived empirically from an organization's tooling and incident investigations.

\section{Agent Training} \label{sec:training}
Training reinforcement learning agents via self-play is challenging in terms of convergence to a globally optimal policy due to the inherently adversarial nature of such training.
To control for potential differences in outcomes related to hyperparameters, we opt to use the same set of hyperparameters across all agents during training.
To this end, we experimented with a number of learning rates and training step sizes.
Our agents were ultimately trained for 3500 steps using an Adam optimizer with a learning rate of $0.0003$ for the actor and $0.0005$ for the critic, with an update batch size of 64. 
For the actor, values of $\{0.00005, 0.0001, 0.0003, 0.0005, 0.001\}$ were tried, and for the critic, values of $\{0.0001, 0.0003, 0.0005, 0.001, 0.0015\}$ were tried, where the critic learning rate was always higher than the actor value for stability.
We also experimented with training step sizes of $[1000, 2000, 3500, 5000, 10000]$ and found that rewards were generally stable within 3000 epochs, even with very low learning rates. 
No other hyperparameters were modified.

\subsection{Multi-Type Training} \label{sec:multitype}
Two of our defender agents, Alternating and Hierarchical, are trained in multi-type scenarios.
In these scenarios, two attacker agents are instantiated from scratch and learn according to their objective -- either the ransomware or APT objective.
During training, one of the two attackers is chosen at random to be the ``active'' attacker, and the defending agent plays against the active attacker using their current policy.
The results are recorded and the agent policy networks are trained over the course of the training run.

In the Alternating case, the defender is a single PPO~\cite{schulman2017proximal} agent that acts against attackers in the game. 
This agent follows the same architecture as the standard PPO agent, with a single actor-critic network learning an action and value policy for deciding what action to take given a particular state. 
By contrast, in the hierarchical case, we leverage HiPPO~\cite{Li2020Sub-policy} and establish a manager network who chooses a subpolicy at some interval $k$.
We consider a ``coarse'' attacker type: a family of attackers $\Phi_{\mathcal{A}} \subseteq \Theta_{\mathcal{A}}$ which may consist of a ``granular'' set of individual $\theta_{\mathcal{A}}$. 
In our case, since $\Theta_{\mathcal{A}}$ consists of only two attacker types, we instantiate only two networks.
For each attacker family, we instantiate a subpolicy network $\pi_{\text{sub}}$.
Thus, at every $k$ timesteps of the game, the manager network assesses the subpolicy which is best performing and chooses actions from that policy for the next $k$ steps, before re-evaluating.
\section{Results}
Our evaluation results demonstrate a number of findings:
\begin{enumerate}
    \item Defenders who see multiple attacker types during training achieve better rewards on average
    \item A simple change in the attacker's condition for victory yields a distinctly different action policy
    \item Learned defender policies have transferability to unseen attacker types
\end{enumerate}

\subsection{Training}
\begin{figure}[h!]
 \begin{subfigure}{0.2\textwidth}
     \includegraphics[width=\textwidth]{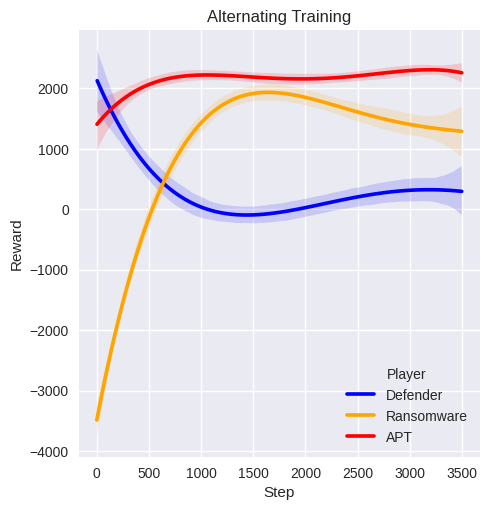}
     \caption{Training time rewards for defender, ransomware, and APT in alternating setting.}
     \label{fig:alt_train}
 \end{subfigure}
 \hfill
 \begin{subfigure}{0.2\textwidth}
     \includegraphics[width=\textwidth]{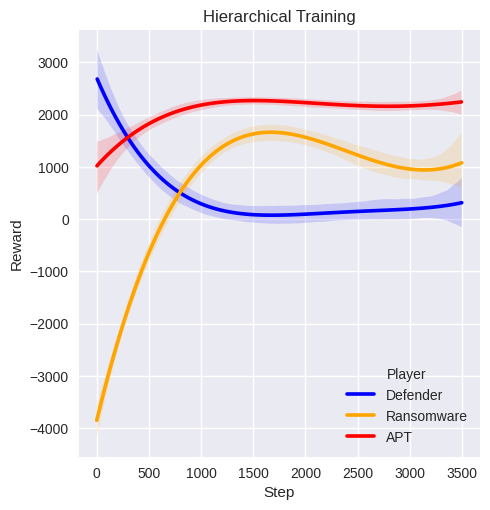}
     \caption{Training time rewards for defender, ransomware, and APT in hierarchical setting}
     \label{fig:hier_train}
 \end{subfigure}
 
 \medskip
 \begin{subfigure}{0.2\textwidth}
     \includegraphics[width=\textwidth]{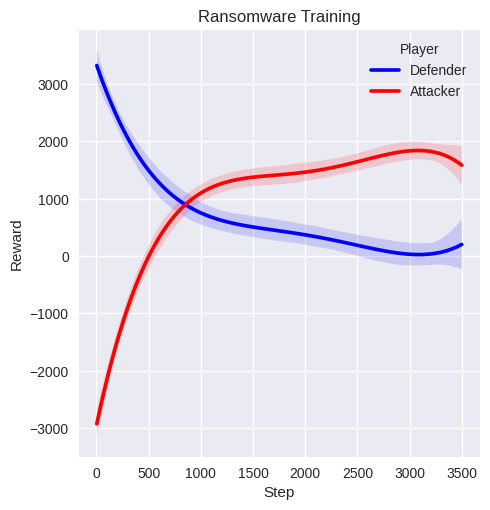}
     \caption{Training time rewards for defender and ransomware attacker}
     \label{fig:ransom_train}
 \end{subfigure}
 \hfill
 \begin{subfigure}{0.2\textwidth}
     \includegraphics[width=\textwidth]{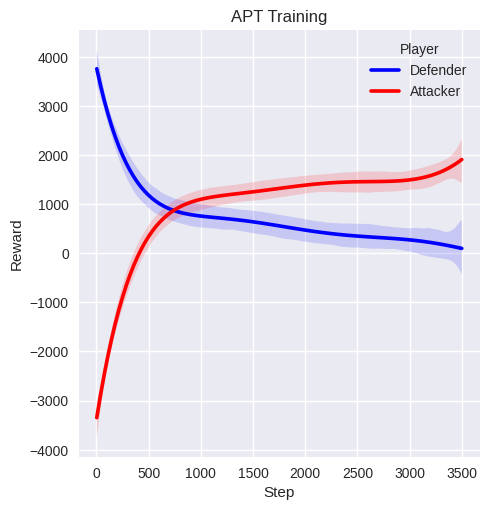}
     \caption{Training time rewards for defender and APT attacker}
     \label{fig:apt_train}
 \end{subfigure}
 \caption{Training curves for all four training settings. Note that curves are smoothed using a best fit line with order 5.}
 \label{fig:training}
\end{figure}

From our training curves in Figure~\ref{fig:training}, we can observe that in all settings, defensive agents initially start out with high levels of reward and converge over time to near-zero reward. 
Although our game is not zero-sum, there is an inverse relationship between attacker and defender rewards.
In the alternating (Figure~\ref{fig:alt_train}) and hierarchical (Figure~\ref{fig:hier_train}) settings, we note that APT rewards are fairly consistent across all training runs, suggesting that more defender adaptation is occurring against the ransomware opponent. 
This is reasonable, considering that the highest scores achieved during training for the defender -- and thus, the learned best policy -- is likely weighted toward ransomware defense.
We also highlight that the combination of training curve smoothing and random attacker assignment overestimates the early returns of the APT model in Figures~\ref{fig:alt_train} and \ref{fig:hier_train}.
In the case of single-type training, pictured in Figures~\ref{fig:ransom_train} and \ref{fig:apt_train}, we note that the reward for learned policies are fairly stable from training step 1500 onward. 
During the course of training, the alternating model played against the APT attacker 1701 times and the ransomware player 1799 times. 
The hierarchical model played against APT 1767 times and ransomware 1733 times.

\subsection{Evaluation}
\begin{figure}[h!]
    \centering
    \includegraphics[width=0.4\textwidth]{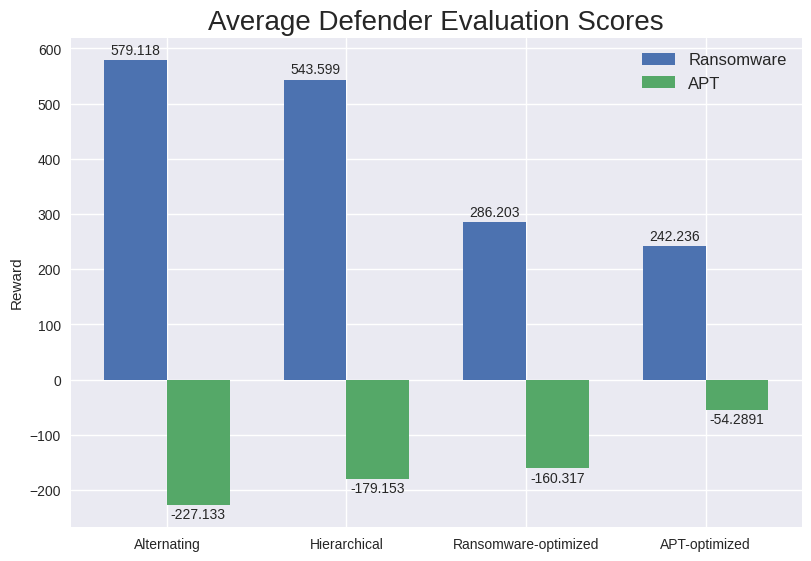}
    \caption{Mean evaluation reward for each defender against ransomware and APT-type attackers}
    \label{fig:mean_eval}
\end{figure}

\begin{figure}[h!]
    \centering
    \includegraphics[width=0.4\textwidth]{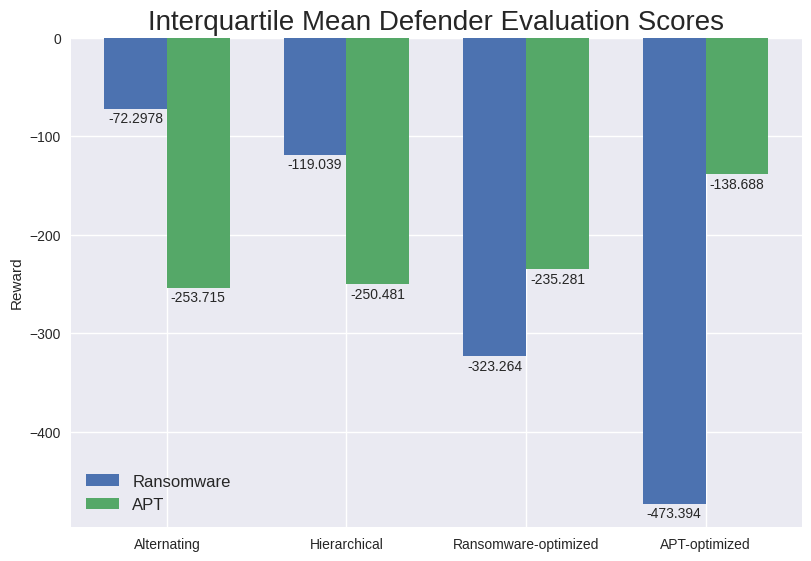}
    \caption{Interquartile mean evaluation reward for each defender against ransomware and APT-type attackers. Note that all rewards in this graph are negative; less negative values are better.}
    \label{fig:iqm_eval}
\end{figure}

Evaluating reinforcement learning findings is a challenge, even more so in our case due to the stochasticity inherent in the environment.
In line with Agarwal \textit{et al.}~\cite{agarwal2021deep}, we consider score distributions and the interquartile mean for our evaluation runs in addition to the mean scores achieved. 
Our evaluation environment is a 50 node network with edges randomly instantiated at each run, ensuring at least 60\% connectivity between nodes and no unconnected nodes. 
The difference in scores between Figure~\ref{fig:mean_eval} and Figure~\ref{fig:iqm_eval} demonstrate the value of examining the interquartile mean. 
While Figure~\ref{fig:mean_eval} shows that the APT-optimized defender does, in fact, perform best on the APT attacker, the hierarchical defender has the best overall average score (383.28) across both attacker types.
On the other hand, Figure~\ref{fig:iqm_eval} demonstrates the truly poor performance of the APT-optimized defender against ransomware attackers and has the defender trained in the alternating setting achieving the best reward (-326.01) across attacker types, with the hierarchical defender performing only marginally worse (-369.52).
In both cases, however, it suggests that the agents who observed both attacker types in training have a meaningful advantage, despite having seen fewer individual instances of each attacker type than either of the specialized models.

\begin{figure}[h!]
    \centering
    \includegraphics[width=0.4\textwidth]{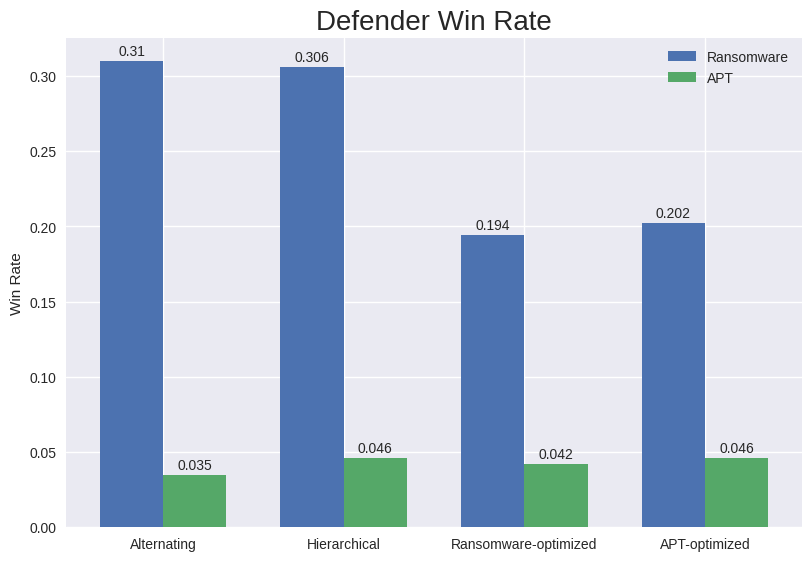}
    \caption{Evaluation win rate for each defender against ransomware and APT-type attackers}
    \label{fig:win_eval}
\end{figure}

Figure~\ref{fig:win_eval} shows the win rates of each defender against both types of attacker.
As the negative values in Figures~\ref{fig:mean_eval} and \ref{fig:iqm_eval} suggest, the attacker achieves their objective in most scenarios, likely due to the ``balanced'' type of the defender being fixed, as mentioned in the section describing defender types.
In fact, the highest win rate of all, the alternating defender against ransomware-type attackers, is only 31\%.
Across all defenders, no single defender manages to achieve better than a 4.6\% win rate versus APT-type attackers.
On average, the hierarchical defender achieves the best win rate (17.6\%), closely tailed by the alternating defender (17.25\%).
We note that these generally disappointing win rates for defenders are, at least in part, due to our assumptions around the detection rate $p$ and false positive rate $q$, and a higher $p$ or lower $q$ would likely yield more impressive outcomes for the defender.

\begin{figure}[h!]
    \centering
    \includegraphics[width=0.4\textwidth]{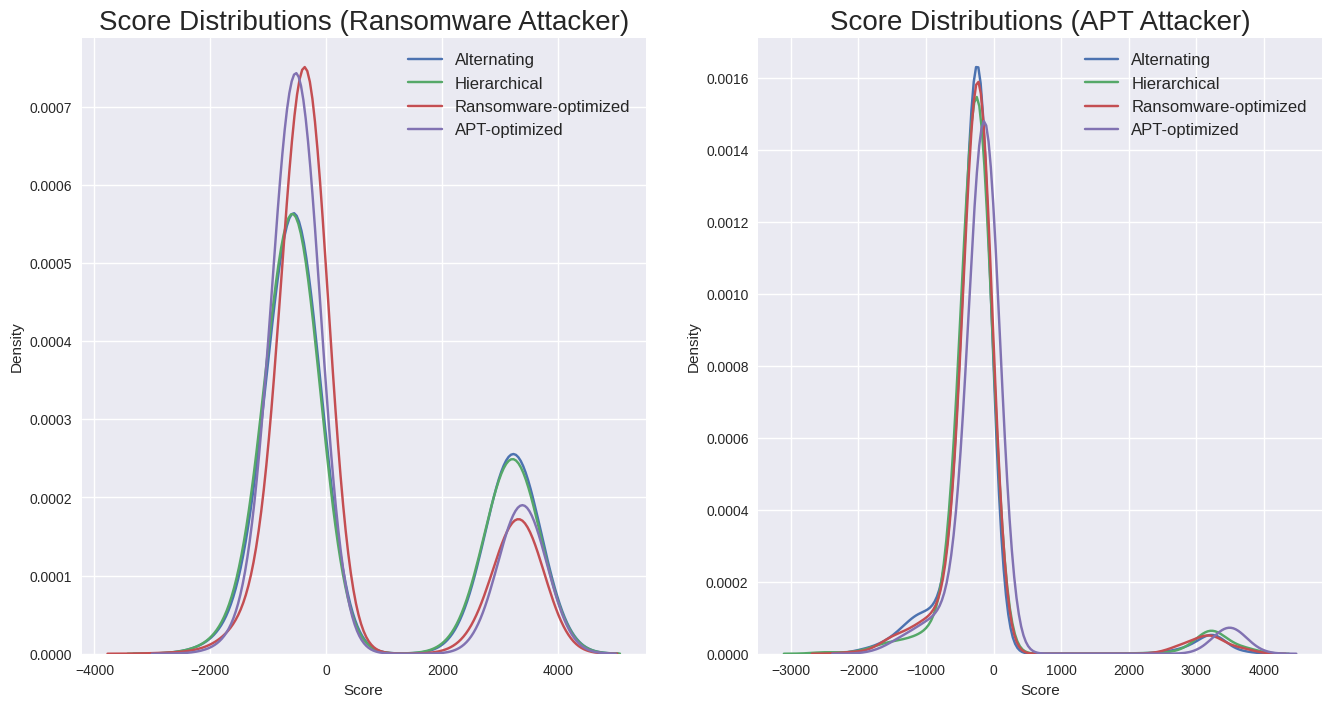}
    \caption{Evaluation score distribution for each defender against ransomware and APT-type attackers. Note the probability density on the y-axes and score values on the x-axes differ between the two charts.}
    \label{fig:score_dist}
\end{figure}

The score distributions in Figure~\ref{fig:score_dist} suggest a number of findings. 
First, across all defenders, the training curves follow a very similar shape against both attacker types and have probability density concentrations at or around the same score value. 
This suggests that the learned policies are fairly similar, though the individual actions and corresponding win rates differ.
Another observation is that although there is transferability of a defender's learned skills -- as evidenced by the APT-optimized defender's performance against ransomware -- the attacker's learned policies are distinct.
As described in our methods section, costs are incurred for actions taken, but there is no penalty other than the cost incurred by actions for losing.
By comparing the two score distribution charts in Figure~\ref{fig:score_dist}, we observe that there is substantially more variance in ransomware outcomes than in APT outcomes, and that APT outcomes tend to have a high density of low-negative losses.
From this, we can infer that APT type attackers, having a much more directed goal of compromising and exfiltrating data from a particular target, are learning a policy that quickly uncovers the target and compromises it, resulting in shorter games with relatively smaller losses. 

\section{Conclusion}
In this work, we have demonstrated that defensive agents trained via reinforcement learning self-play are capable of learning strategies to mitigate multiple types of attackers.
In particular, we have shown that agents trained in multi-type settings -- our ``alternating'' and ``hierarchical'' agents -- perform meaningfully better on average than agents trained to specialize against one particular attacker type. 
We highlight that as additional attacker types are introduced, the hierarchical agent has an advantage for scaling in needing only to learn the high-level policy.
Although we leverage HiPPO~\cite{Li2020Sub-policy} in this work, online algorithms for mixtures of experts like exponential weighting~\cite{littlestone1994weighted} potentially offer a major advantage in quickly adapting to novel threats.
We wish to explore the pros and cons of this sort of online learning methodology in future work.

One limitation of this work is that YAWNING-TITAN's network node states restrict the available attacker and defender action spaces.
Although the policies learned by our two attacker types do differ in a meaningful way, this limitation makes it difficult to introduce additional attacker types who may differ in the actual techniques used. 
In future work, we seek to apply our findings to more fully-featured simulated environments with richer states and actions such that we can emulate a larger number of adversaries and apply our findings in real-world environments.

\bibliography{references}

\end{document}